\documentstyle[12pt,psfig,epsf]{article}
\newcommand{\be}{\begin{equation}}
\newcommand{\ee}{\end{equation}}
\newcommand{\la}{\langle}
\newcommand{\ra}{\rangle}
\begin{document}
\hfill  NBI-HE-98-31 

\hfill  NRCPS-HE-98-13

\hfill  MPS-RR-98-26

\hfill  NTUA-98/99

\vspace{24pt}
%\begin{titlepage}
%\title{
\begin{center}
{\large \bf Four-dimensional gonihedric gauge spin system}%title ends

\vspace{24pt}
%\author{ 
{\sl J.Ambj\o rn}\\
The Niels Bohr Institute, University of Copenhagen, \\
Blegdamsvej 17, DK-2100 Copenhagen \O , Denmark\\
{\tt email:ambjorn@nbi.dk}

\vspace{1cm}

{\sl G.Koutsoumbas}\\
Physics Department, National Technical University, \\
Zografou Campus, 15780 Athens, Greece\\
{\tt email:kutsubas@central.ntua.gr}
\vspace{1cm}

{\sl G.K.Savvidy}

National Research Center Demokritos,\\
Ag. Paraskevi, GR-15310 Athens, Greece \\
{\tt email:savvidy@argo.nrcps.ariadne-t.gr}
%}%author ends
%}
%\date{}%in order NOT to write the date
%\maketitle
\end{center}
\vspace{60pt}

\centerline{{\bf Abstract}}

\vspace{12pt}
\noindent
We perform Monte Carlo simulations of a four-dimensional gauge invariant spin
system which describes random surfaces with gonihedric action.  
We develop the analogy 
between the flat-crumpled phase transition of the lattice surface model and
the liquid-gas phase transition of non-ideal gases,
and identify the self-intersection coupling constant $k$ of the 
surface model with 
the pressure $P$. As $k$ increases the system moves to 
a critical point in complete 
analogy with the situation for non-ideal gases, where 
the liquid and the gas phases 
approach  each other with increasing  $P$. 
We measure vacuum expectation values of various operators 
and the corresponding critical indices.

%\end{abstract}
%\thispagestyle{empty}
%\end{titlepage}

\newpage

\pagestyle{plain}
%\pagenumbering{roman}
\section{Introduction}

%\vspace{.5cm}

In this article we  consider a model of two-dimensional random 
surfaces embedded into an Euclidean lattice $Z^4,$ where a closed surface 
is associated with a collection of 
plaquettes. The surfaces may have self-intersections in the form of four and 
six plaquettes intersecting on a link. The edges of the surface with 
self-intersections comprise 
the {\it singular part} of the surface. The edges of the surface where only two 
plaquettes are intersecting comprise the {\it regular part} of the surface. 

Various  models of  random surfaces built out of plaquettes have been
considered in the literature \cite{weingarten}. In this article 
we  consider the so-called gonihedric model which 
has been defined in refs.\ \cite{sav1,sav} and we compare it with 
the model with area action \cite{weg,bal}. 
The gas of random surfaces defined in 
\cite{weg} corresponds to the partition 
function with Boltzmann weights proportional to the total number $n$ 
of plaquettes, that is to Nambu-Goto area functional.
The gonihedric model of random surfaces will correspond to a statistical system 
with weights proportional to the total number $n_2$ of non-flat edges of the 
surface,
that is to a linear size of the surface \cite{sav1}. 
The edges are non-flat when 
the dihedral angle between plaquettes is not equal to $\pi$. The weights 
associated with self-intersections are proportional to $k n_4$ and $k n_6$ 
where $n_4$ and $n_6$ are the number of edges with four and six intersecting 
plaquettes, and $k$ is the self-intersection coupling constant \cite{sav1,sav}.

To study the statistical and scaling properties 
of the system one can directly 
simulate surfaces by gluing together plaquettes following the rules described 
above, that is on every link only an even number of plaquettes should intersect 
and the weights are proportional to the area $n$ or gonihedricity 
$n_2 + 4 k n_4 + 12 k n_6$. But it is much 
easier to use the duality between random surfaces and spin systems on 
Euclidean lattice. This duality allows us to  study  equivalent spin 
systems with specially adjusted interaction between spins \cite{sav}.

In four dimensions a spin Hamiltonian which is 
equivalent to an area action, is well 
known. It represents a gauge invariant 
spin system with  one-plaquette interaction 
terms \cite{weg}
\be
H^{4d}_{area}= -\frac{1}{g^{2}} \sum_{\{plaquettes\}}
(\sigma\sigma\sigma\sigma) ,                                 \label{area}
\ee
where spins are located on the links and the summation is over all 
plaquettes on the lattice. Likewise there exists a gauge invariant 
Hamiltonian which is equivalent to the 
gonihedric action. It has a more complicated interaction 
between spins  \cite{sav}:
$$
H^{4d}_{gonihedric}= -\frac{5\kappa-1}{g^{2}} \sum_{\{plaquettes\}}
(\sigma\sigma\sigma\sigma) + \frac{\kappa}{4g^{2}} 
\sum_{\{right~angle~plaquettes\}}
(\sigma\sigma\sigma\sigma_{\alpha})^{rt}
(\sigma_{\alpha}\sigma\sigma\sigma)$$
\be
-\frac{1-\kappa}{8g^{2}} 
\sum_{\{triples~of~right~angle~plaquettes\}}
(\sigma\sigma\sigma\sigma_{\alpha})^{rt}
(\sigma_{\alpha}\sigma\sigma\sigma_{\beta})^{rt}
(\sigma_{\beta}\sigma\sigma\sigma)  ,               \label{gonih}
\ee
where the spins interact within a three-dimensional cube of the four-dimensional 
lattice. The coupling constant $k$ monitors the interaction 
at the singular parts of the surface where self-intersections take place. If 
$k$ is equal to zero then there is no repulsion associated with 
self-intersections, and if $k$ is large the surfaces are self-avoiding 
\cite{sav}.

The partition functions for both systems have the form 
$$
Z(\beta) = \sum_{ \{\sigma \} } e^{-\beta~g^{2} H/4}, 
$$
where the summation is over all spin configurations. 
This partition function can be 
represented in a dual form as a sum over two-dimensional surfaces of the 
type described above, embedded into a four-dimensional lattice \cite{sav1,sav}:
\be
Z(\beta) = \sum_{\{surfaces~M\}} e^{-\beta~\epsilon(M)}  \label{partfan}
\ee
where $\epsilon(M)$ is the energy of the surface $M$ constructed from plaquettes
such that each edge of the surface is contained  in either two, four or six 
plaquettes. Those edges of the surface where four and six plaquettes intersect
comprise the curves of self-intersections.

If a two-dimensional surface $M$ has $n$ plaquettes, $n_{2}$ edges with 
two-plaquette intersection, $n_{4}$ edges with four-plaquette 
intersection (~more precisely there are two different geometries of 
four-plaquette 
intersections, the corresponding numbers are equal to $\bar{n}_{4}$ and 
$\bar{\bar{n}}_{4}$ \cite{sav}~) and
$n_{6}$ edges with six-plaquette intersection, then the total energy 
$\epsilon(M)$ of the surface $M$ 
with Hamiltonian (\ref{area}) is equal to $n$: 
$$
\epsilon_{area}(M) = n(M).
$$
For the gonihedric system with Hamiltonian (\ref{gonih}) the energy 
is related to the number of non-flat edges of $M$ by:
\be
\epsilon_{gonihedric}(M) = n_{2}(M) + 4k \bar{n}_{4}(M) + 
(6k-1) \bar{\bar{n}}_{4}(M) + 12k n_{6}(M)  .              \label{energyk}
\ee
If $k$ is small self-intersections are permitted, 
but when $k$ increases there is 
strong repulsion along the curves of self-intersections, and the surfaces tend 
to be  self-avoiding \cite{sav}. 

The total energy is defined as 
$$
E = -\partial{ln Z}/\partial{\beta}=
\la n_{2} + 4 k \bar{n}_{4} + (6 k-1)\bar{\bar{n}}_{4} + 12k n_{6} \ra_{c}
$$ 
and specific heat as 
\be
C = \beta^{2}\frac{\partial^{2}{ln Z}}{\partial{\beta^{2}}} =
-\beta^{2} \frac{\partial{E}}{\partial{\beta}} .   \label{specif}
\ee
The derivative of the partition function (\ref{partfan})
with respect to $k$ gives an {\it average length} $\mu$ of the 
self-intersection curves
\be
\mu = -\frac{1}{\beta}\frac{\partial{ln Z}}{\partial{k}} =
\la 4 \bar{n}_{4} + 6 \bar{\bar{n}}_{4} + 12 n_{6}\ra_{c}   \label{orderpar}
\ee
which we suggest to consider as a {\it disorder parameter}, because it is zero 
in the
low-temperature phase and is nonzero at high-temperature phase  \cite{cut}. The 
second derivative with respect to $k$ 
defines the {\it intersection susceptibility}
\be
\chi=   \frac{1}{\beta^{2}} \frac{\partial^{2}{ln Z}}{\partial{k^{2}}} =
 -\frac{1}{\beta}\frac{\partial{\mu}}{\partial{k}} =
\la (4 \bar{n}_{4} + 6 \bar{\bar{n}}_{4} + 12 n_{6})^2\ra_{c} - 
\la 4 \bar{n}_{4} + 6 \bar{\bar{n}}_{4} + 12 n_{6}\ra^{2}_{c}.   \label{suscept}
\ee
The average area $S$ and the area susceptibility $\chi^S$ are defined 
analogously:
$$
S = \la n \ra_{c},~~~~~~\chi^{S}=\la n^2 \ra_{c} - \la n \ra^{2}_{c}.
$$
All these observables can be expressed 
in terms of spin variables and 
then used in Monte Carlo simulations \cite{sav,cut}.

\begin{figure}
\centerline{\hbox{\psfig{figure=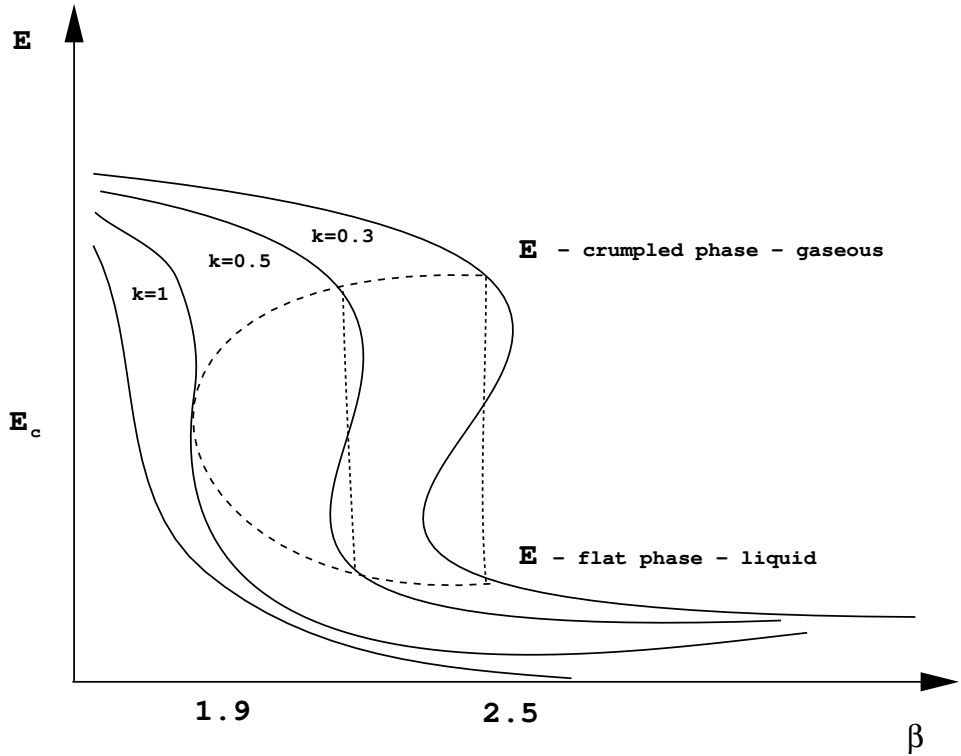,height=10cm,angle=00}}}
\caption[fig1]{Schematic exposition of the equation of state for the 
4D-system (\ref{gonih}), which is similar to  Van der Waals-like curves and 
coexistence curve for the real gases. 
The system was heated and then cooled in small $\beta$ steps \cite{cut}.}
\label{fig1}
\end{figure}

In \cite{cut} it was demonstrated that the system, with a sufficiently 
large value for the 
self-intersection coupling constant $k$ ($k_{c}\approx 1$), undergoes a 
second-order phase transition at some critical temperature $\beta_{c}$.
This result suggests the existence of continuum field theory in four 
dimensions and our aim is to study further the phase structure of the system, 
the scaling properties of different operators,  and to compute  
the corresponding critical indices. For this purpose it is helpful to develop 
the analogy between flat-crumpled phase transition of the gonihedric surface 
model \cite{sav,cut} and liquid-gas phase transition of non-ideal gases, and to 
identify the intersection coupling constant $k$ 
with the pressure $P$, as well as the intersection 
susceptibility (\ref{suscept}) with compressibility $K_{T}$ \cite{fisher}.

\section{Analogy with non-ideal gases}

Before presenting new  results for gonihedric system 
let us  describe
the known behaviour of spin systems with an area-type action \cite{creutz1}.
As mentioned in the introduction, the random surfaces with an area action can 
be represented by the 
gauge invariant Hamiltonian (\ref{area}) \cite{weg}. 
In four dimensions this gauge invariant spin system 
is self-dual and the critical temperature is equal to 
$\beta_{c} = \frac{1}{2} ln(1+\sqrt{2})$ \cite{weg}. 
Monte Carlo simulations strongly suggest that the phase transition in 
4d $Z_2$ gauge invariant 
spin system is of  first order \cite{creutz1,creutz2,creutz}. A clear 
signal of the first order nature is obtained 
by measuring the one-plaquette average  in a  thermal 
cycle. 

A thermal cycle
of the statistical system  provides a general overview of its phase structure 
and is very helpful in defining the regions of the phase transitions 
\cite{creutz1}, and pronounced hysteresis 
is strongly indicative of a first-order phase transition \cite{creutz1}. 
A typical hysteresis curve will look like the curve  
corresponding to $k=0.3$ on Fig.\ 1, and is due to the 
meta-stability of the ordered phase at high temperature and of the
disordered phase at low temperature.

The phase structure of the self-avoiding 
random surface model in four dimensions, 
with the gonihedric action (\ref{gonih}),
has  recently been analysed in \cite{cut}. 
The Monte Carlo simulations of the 
corresponding gauge invariant spin system demonstrate that the critical 
behaviour of the system essentially depends on the self-intersection 
coupling constant $k$. For small values of $k$ a thermal cycle shows a clear
hysteresis loop, thus favouring a first order phase transition 
(see  the Van der Waals-like curves k=0.3 and k=0.5 on Figure 1). Two 
distinct stable phases appear at the critical temperature with large 
differences between their energy densities: 
\be
\delta E =  E_{crumpled} - E_{flat},
\ee
(where the names {\it crumpled} and {\it flat} refer to the 
typical embedding of the surfaces on the hyper-cubic lattice),
see Figure 1 and Figures 1-9 in Ref.\cite{cut}. 
This first order phase transition
at small $k$ is very similar to the first order phase transition in the model
with the one-plaquette  area action (\ref{area}).

We have to define also a gap in the intersecting energy $\mu$ 
\be
\delta \mu = \mu_{crumpled} - \mu_{flat},
\ee
and a gaps in the simple bendings $n_{2}$ and the average area $S$ as 
\be
\delta n_{2} = n_{2~crumpled} - n_{2~flat},~~~~~~~~~\delta S = S_{crumpled} 
- S_{flat}
\ee

As the self-intersection coupling constant $k$ increases, the  energy
difference $\delta E$~--~the gap between the densities $E_{crumpled}$ and 
$E_{flat}$ of coexisting phases~--~ tends continuously to zero 
(see Figure 2 and Table 1). The same is true for the gaps in the cases of
intersecting energy, of simple bendings and of average area (see Figure 2 
and Table 1). These gaps  tend to zero  also when the volume increases
for the fixed value of the coupling constant $k_{c} \approx 1$, as it can be 
seen from Table 1.

\newpage

\begin{center}
\begin{tabular}{c}
\hline
\hline
TABLE 1\\
\hline
\hline
\\
Volume dependence of the total energy and intersection energy:\\
\end{tabular}
\end{center}

\begin{tabular}{||c|c|c|c|c|c|c||}
\hline
\hline
k&Volume&$\beta_{crit}$&$C_{max}/V$&$\delta E$ &
$\chi_{max}/V$&$\delta \mu$\\
\hline
\hline
0.3&$6^4$&2.64-2.66&0.101(3)&0.64(2)&0.0133(6)&0.23(1) \\
\hline
0.3&$10^4$&2.50-2.55&0.108(5)&0.66(3)&0.015(1)&0.25(2) \\
\hline
0.3&$12^4$&2.50-2.54&0.10(1)&0.64(6)&0.014(2)&0.24(3) \\
\hline
0.3&$20^4$&2.50-2.54&0.1050(8)&0.648(5)&0.01471(8)&0.243(1) \\
\hline
\hline
1.0&$6^4$&1.840-1.849&0.018(5)&0.26(6)&0.00021(7)&0.003(1) \\
\hline
1.0&$8^4$&1.904-1.906&0.015(2)&0.24(3)&0.00010(2)&0.0019(5) \\
\hline
1.0&$10^4$&1.919-1.920&0.011(2)&0.20(3)&0.00007(1)&0.0017(3) \\
\hline
1.0&$12^4$&1.920-1.921&0.009(2)&0.18(4)&0.00007(1)&0.0016(2) \\
\hline
1.0&$16^4$&1.915-1.920&0.009(1)&0.19(2)&0.00006(1)&0.0016(2) \\
\hline
1.0&$20^4$&1.925-1.926&0.008(2)&0.17(4)&0.00006(1)&0.0015(3) \\
\hline
\hline
\end{tabular}

\begin{center}
\begin{tabular}{c}
\\
\\
Volume dependence of the simple bendings and the plaquettes:\\
\end{tabular}
\end{center}

\begin{tabular}{||c|c|c|c|c|c|c||}
\hline
\hline
k&Volume&$\beta_{crit}$&$\chi^{n_{2}}_{max}/V$&$\delta n_{2}$
&$\chi^{S}_{max}/V$&$\delta S$\\
\hline
\hline
0.3&$6^4$&2.64-2.66&0.041(2)&0.41(1)&0.08(3)&0.6(2) \\
\hline
0.3&$10^4$&2.50-2.55&0.042(2)&0.41(1)&0.145(4)&0.76(2) \\
\hline
0.3&$12^4$&2.50-2.54&0.041(3)&0.41(2)&0.146(9)&0.76(5) \\
\hline
0.3&$20^4$&2.50-2.54&0.0405(4)&0.405(4)&0.141(1)&0.754(5) \\
\hline
\hline
1.0&$6^4$&1.840-1.849&0.018(4)&0.26(6)&0.005(2)&0.15(4) \\
\hline
1.0&$8^4$&1.904-1.906&0.013(2)&0.22(3)&0.007(2)&0.16(3) \\
\hline
1.0&$10^4$&1.919-1.920&0.009(1)&0.19(3)&0.008(1)&0.15(2) \\
\hline
1.0&$12^4$&1.920-1.921&0.0075(7)&0.17(2)&0.0085(8)&0.16(2) \\
\hline
1.0&$16^4$&1.915-1.920&0.008(1)&0.18(2)&0.010(2)&0.20(3) \\
\hline
1.0&$20^4$&1.925-1.926&0.007(1)&0.16(3)&0.008(2)&0.17(4) \\
\hline
\hline
\end{tabular}

\vspace{1.0cm}

The first order phase 
transition becomes weaker and then disappears. The limiting density 
$E_{c}$ and the corresponding temperature $\beta_{c}$ and the coupling constant 
$k_{c}$ define the critical point of a second-order phase transition.
Thus the intersection coupling constant $k$ plays an important role and 
by monitoring $k$ one can drive
the system to a critical point where there might be a second order transition.

\begin{figure}
\centerline{\hbox{\psfig{figure=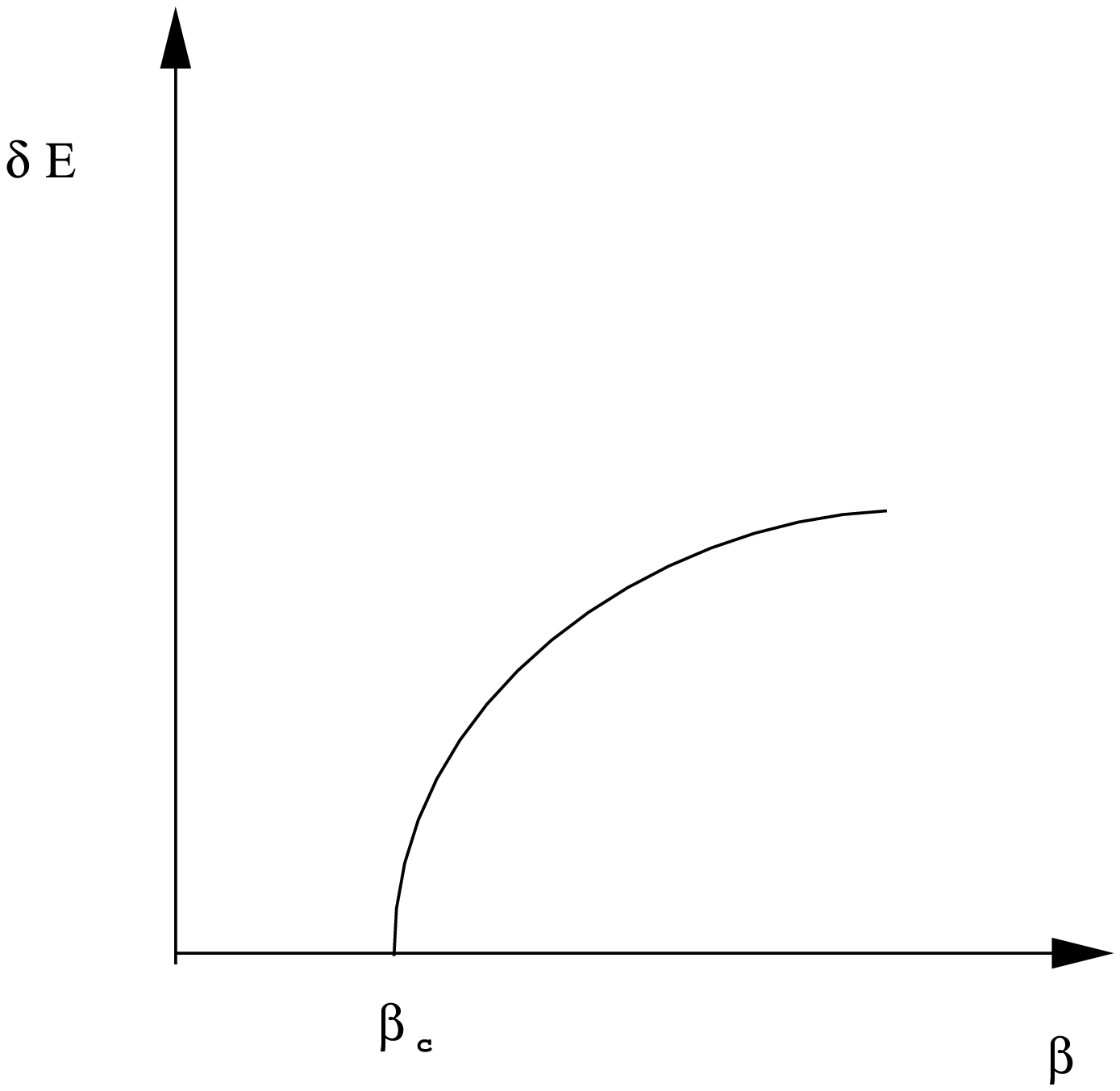,height=6cm,angle=00}}}
\caption[fig2]{Coexistence curve: plot of the gap $\delta E  = E_{crumpled}$ 
- $E_{flat}$ vs $(\beta - \beta_{c})$.} The gap vanishes as
$(\beta - \beta_{c})^{1-\alpha}$
\label{fig2}
\end{figure}

This picture of the phase transition of the four-dimensional system has a great 
similarity with the liquid-gas critical point in three dimensions, and  
the coexistence curves on 
Fig.\ 1 and 2 is the analogue 
of the coexistence curve of the liquid and the gas 
phases. We will associate the flat phase of the low temperature gauge system 
with the liquid phase and the crumpled high temperature phase with the 
gaseous phase. In accordance with the proposed analogy, the self-intersection 
coupling constant $k$ should be associated with the pressure P. 

Indeed, the 
non-ideal gas condenses to a liquid state by cooling at fixed pressure.
As pressure increases condensation takes place at higher temperatures and two 
phases merge at the critical point. The same phenomenon takes
place in the gonihedric
system in four-dimensions. By heating the low temperature 
flat-liquid phase at fixed $k$ we ``evaporate'' clouds of crumpled surfaces 
(see Figure 1) at some temperature $\beta_{k}$. As the self-intersection 
coupling constant $k$ increases evaporation of the clouds takes place at 
higher and higher  
temperatures (see Figure 1 and Table 1) 
and the energy gap between two phases decreases 
and tends to zero (see Figure 2). Thus the self-intersection coupling constant
$k$ allows us to move the system from one 
Van der Waals-like curve to another and finally to the critical point.
In the case of the area-action
the system does not have such an additional coupling constant 
and is described by single Van der Waals-like curve \cite{creutz1}.

The fact that the clouds of evaporated surfaces 
are indeed crumpled can be seen from the self-intersection energy density 
$\epsilon_{intersection}$.
This part of the total energy $\epsilon_{gonihedric}$ is equal to zero at 
the low temperature flat phase, and jumps to a nonzero 
value at the evaporation point $\beta_{k}$ \cite{cut}. 
As $k$ increases the evaporation takes place 
at higher temperatures $\beta_{k_1} < \beta_{k_2}$ ~~($k_{2}  < k_{1}$) and 
two phases merge at the critical point $(\beta_c,k_c)\approx (1.925,1)$.
The curve of the 
first-phase transitions in the $(\beta , k)$ coupling constant plane 
ends at the critical point $(\beta_{c}, k_{c})$ as illustrated in Fig.\ 3.

\begin{figure}
\centerline{\hbox{\psfig{figure=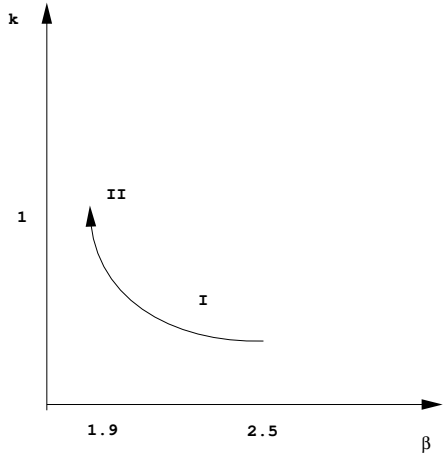,height=6cm,angle=00}}}
\caption[fig3]{The curve of the 
first-order phase transitions in the $(\beta , k)$ coupling constant plane 
is ended at the critical point $(\beta_{c}, k_{c})$.}
\label{fig3}
\end{figure}

\section{Monte Carlo simulation of gonihedric system}

Before describing the results, let us briefly explain some points 
having to do with the simulation. One of the basic issues is the
order of the phase transition, which has to be determined by the 
study of the volume dependence of several quantities, namely the 
peaks of the  variances $\Delta Q^2 \equiv \la Q^2 \ra-\la Q \ra^2$ 
(specific heat C, intersection susceptibility $\chi$, area 
susceptibility $\chi_{S}$) of the measured 
quantities $Q$ (total energy E, average length $\mu$, area $S$). 
We recall that for a first order transition 
the peak divided by the volume $\delta Q^2|_{max} \equiv 
\frac{\Delta Q^2|_{max}}{V}$ 
should be volume independent; on the other hand, 
for a continuous transition, the volume dependence will be weaker
and should be characterized by critical exponents: $\alpha/d\nu$
for the specific heat C,  $\gamma/d\nu$
for the intersection susceptibility and $\tilde{\gamma}/d\nu$
for the area susceptibility $\chi_{S}$ 
\be
\Delta Q^2|_{max} \approx V^{index} . \label{variance}
\ee
A quantity related to the variance is the so called gap. While simulating
the system we found out that if we average over only one of the 
meta-stable states, the resulting values $\la Q \ra_{crumpled},
\la Q \ra_{flat}$ 
(for the quantity Q, say) were well separated; the
intermediate values have rarely shown up. One may define the gap $g(Q)$ as
the difference: $g(Q) \equiv \la Q \ra_{crumpled}-\la Q \ra_{flat}.$ 
It is easy to show that, in the 
situation just described, $g(Q)$ is related to the peak of the variance through
$\delta Q^2|_{max} \approx \frac{g^2(Q)}{4}.$ 
Thus, if the phase transition is of
first order, the lack of volume dependence for $\delta Q^2|_{max}$ will result
to a volume independence of  $g(Q)$ as well. Thus, the volume 
independence of the gap may serve as an additional indicator of the
order of the phase transition. We note that for a continuous phase 
transition the gap will decrease for increasing volume.

We have performed  Monte Carlo simulations at several
values of $k$ and $\beta$ close to the phase transition lines
that have been determined in \cite{cut}. We choose to show in the 
sequel the results for $k=0.3$ and $k=1.0$ 
for the lattice sizes $6^4,~8^4,~10^4,~12^4,~16^4,~20^4.$
The observables we have measured are the following:
\begin{itemize}
\item  The total energy $H^{4d}_{gonihedric},$ as defined in equation 
(\ref{gonih}) and  (\ref{energyk}).
\item The part of the energy related to the simple bendings
of plaquettes ($n_{2}$).
\item The intersection energy, which is defined as the difference of the two
previous quantities (\ref{orderpar}) .
\item The plaquette ($n$).
\end{itemize}

Table 1 displays the results of the measurements. For $k$=0.3 or $k$=1.0 the 
critical values for $\beta$ are reported for the various lattice volumes
together with the peaks of the variances and the energy gaps
(uncertainty in the last digit is put 
in parentheses).  In the upper half of the table one finds the results for
the total energy and for the interaction energy; 
in the lower half one finds the the results for 
the simple bendings and for the plaquette. Let us first observe that for 
$k$=0.3 the gap and the quantities $\delta Q^2|_{max}$ are volume 
independent within  errors, indicating that for this $k$ 
the phase transition is of first order; one may also check that the 
relation $\delta Q^2|_{max} \approx \frac{g^2(Q)}{4}$ is 
approximately satisfied. For $k$=0.3 the total energy and the simple 
bendings have a sizeable difference which shows up as intersection energy.
We should remark that the volume independence of the gap and the
peak of the variance also holds for 
$k$=1.0 in the special case of the plaquette.

Having found the relevant $\Delta Q^2|_{max}$ we can extract 
the critical exponents using (\ref{variance}).

\begin{center}
\begin{tabular}{c}
\hline
\hline
TABLE 2   Coefficients a and b of the best fit a+b x  \\
\hline
\hline
\end{tabular}
\end{center}
\begin{center}
\begin{tabular}{||c|c|c|c|c|c||}
\hline
\hline
k   &       &a&$\delta$ a&b&$\delta$ b \\
\hline\
0.3 &Energy &-2.34&0.07&1.008&0.006 \\
0.3 &Intersections &-4.4&0.1&1.02&0.01 \\
0.3 &Bendings &-3.16&0.08&0.997&0.007 \\
0.3 &Plaquette &-1.88&0.09 &0.993&0.007 \\
\hline\
1.0 &Energy &-2.9&0.5&0.83&0.05 \\
1.0 &Intersections &-7.9&0.6&0.83&0.06 \\
1.0 &Bendings &-2.9&0.4&0.82&0.04 \\
1.0 &Plaquette &-5.8&0.5  &1.11&0.05\\
\hline
\hline
\end{tabular}
\end{center}

The results are depicted in Fig.\ 4. The values of $\log(\Delta Q^2|_{max})$
are plotted versus the logarithm 
of the volume for the four measured  quantities. Fig.\ 4 also contains 
the best straight line fits. The best parameters (with their errors)
are collected in Table 2. It is striking that for this small $k$ the slope 
is equal to one within the errors. This means that the variance is 
proportional to the volume, that is the ratio  
$\delta Q^2|_{max} \equiv \frac{\Delta Q^2|_{max}}{V}$ is a constant, 
which is once more the sign of a first order phase transition.

\begin{figure}
\centerline{\hbox{\psfig{figure=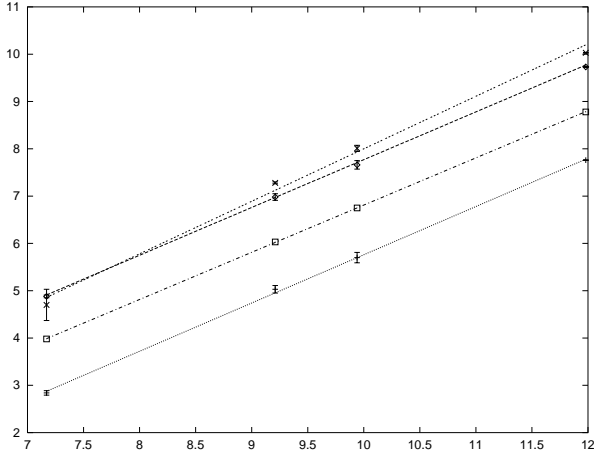,height=6cm,angle=-90}}}
\caption[fig4]{The critical indices at k=0.3. All quantities scales with the
lattice volume as a first power. Intersections are represented by crosses, 
bendings by boxes, plaquettes by diamonds and energy by x's.}
\label{fig4}
\end{figure}

\begin{figure}
\centerline{\hbox{\psfig{figure=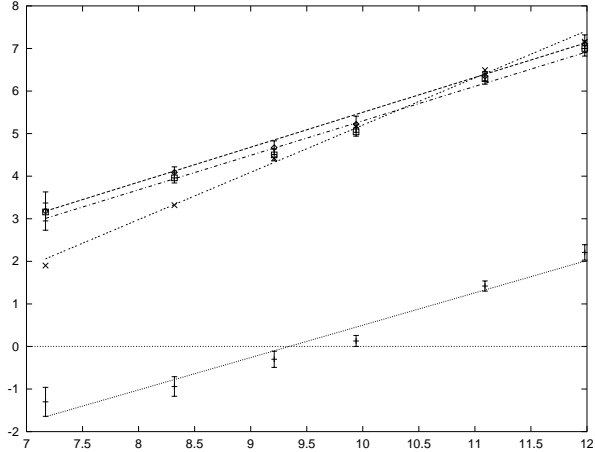,height=6cm,angle=-90}}}
\caption[fig5]{The critical  indices at k=1.}
\label{fig5}
\end{figure}

For $k$=1.0 both the gap and 
$\delta Q^2|_{max}$ decrease with increasing volume, thus indicating
that the transition 
has become second order. One should also notice that for k=0.3 the total 
energy and the bendings have a sizeable difference, which shows up
as intersection energy; on the contrary, for $k$=1.0 the energy 
is almost entirely built up from simple bendings and plaquettes, 
which is reflected to the very small value of the intersection energy. 
One should also notice the quite large critical exponent for the plaquette.

To substantiate these claims, we plot in Figure 5 $Log(\Delta Q^2|_{max})$
versus the logarithm of the volume and determine the relevant critical 
exponent. We find that the best slopes are around 0.83 for all 
quantities with the exception of the plaquette.
The small value of the exponent, as compared against 1.0 above, also 
lends support to the view that the transition is continuous.
 
Thus we conclude that finite size scaling favours a second order 
phase transition at $(\beta_c,k_c)$ with seemingly nontrivial exponents
(but see next section), 
and it is apparent that the transition 
has great qualitative similarity with the liquid-gas second order 
transition at the coexistence point $(\beta_c,P_c)$.

\section{Discussion}

We have analyzed numerically the four-dimensional 
gonihedric gauge spin system. In its random surface 
representation it exhibits a flat-crumpled transition
which is in  close analogy with the liquid-gas transition
of non-ideal gases, and with the 
intersection coupling constant $k$ of the spin system 
playing the role of the pressure $P$. We found evidence 
that in the neighborhood of $k=1.0$ the first order transition
between the flat and crumpled phase changes to a second order
transition. Again, this is in accordance with the liquid-gas picture.
We  found $\alpha/d\nu = 0.83\pm 0.06$ (except for the plaquette energy
which has $\alpha/d\nu \approx 1.11$). It could indicate a non-trivial 
scaling and thus a new four-dimensional class of critical systems.
However, before jumping to these conclusions one should keep in mind that 
it is notoriously difficult to measure the exponents at a point where the 
first order transition stops and changes to a higher order 
transition. The true exponents tend to be masked by so-called 
pseudo-critical exponents of the first order transition: $\nu =1/d$,
where $d$ is the dimension of space \cite{berker}. 
In our case it would mean that 
a pseudo-critical $\nu$ should be 0.25 if we did measure the 
first order pseudo-critical exponent. From the measured values
of $\alpha/d\nu$ and assuming hyper-scaling ($2-\alpha = \nu d$)
we get $\nu = 0.27\pm 0.01$ for $\Delta Q^2$ coming from energy, 
intersections and bendings.
These values seem uncomfortable close to $\nu = 0.25$ and although 
the measured value of $\alpha/d\nu$ is more that two standard deviations
away from the first-order value $\alpha/d\nu =1$, systematic errors 
might be larger than anticipated. Thus we cannot rule out that we see 
a very slow shift away from this value and towards the true value for the 
second order transition, even if we have carefully checked for 
finite volume dependence. If this is the case the true 
value of the critical exponent is of course unknown. 
Further studies seem necessary in order 
to settle this interesting question in an unambigous way.

\section{Acknowledgement} One of the authors (G.K.S) was supported in part 
by the EEC Grant no. ERBFMBICT972402. J.A. acknowledges the support 
from MaPhySto -- Centre for Mathematical Physics and Stochastics --
which is financed by Danish National Research Foundation. G.K. acknowledges 
the support from EEC Grant no. ERBFMRX CT 970122.

\vfill
\end{document}